\documentclass{svjour3}                     
\usepackage{amsmath}
\usepackage{amsfonts}
\usepackage{amssymb}
\usepackage{latexsym,epsfig}
\usepackage{graphicx}
\journalname{bjp}
\begin{document}

\title{On a $\chi^2$-function with previously estimated background \thanks{This work was partially supported by National Council for Scientific and Technological Development (CNPq) under grant 402846/2016-8}
}


\author{Fernando M.L. Almeida Jr. \and   A.A.Nepomuceno
}

\institute{Fernando M.L. Almeida Jr. \at
              Depto. de Ciencias Naturais, Universidade Federal de S\~ao Jo\~ao del Rei, S\~ao Jo\~ao del Rei, BR-36301-160 MG Brazil \\
              \email{fmarroquim@ufsj.edu.br}           
           \and
           A.A.Nepomuceno \at
           Departamento de Ciencias da Natureza, Universidade Federal Fluminense,
           Rio das Ostras, RJ, Brazil. \\
           \email{asevedo@gmail.com}
}

\date{Received: date / Accepted: date}
\maketitle

\begin{abstract}
There are intensive efforts searching for new phenomena in many present and future scientific experiments such as  LHC at CERN, CLIC, ILC and many others. These new signals are usually rare and frequently contaminated by many different background events. Starting from the  concept of profile likelihood we obtain what can be called a profile $\chi^2$-function for counting experiments which has no background parameters to be fitted.  Signal and background statistical fluctuations are automatically taking in account even when the content of some bins are zero. This paper analyzes the profile $\chi^2$-function for fitting binned data in counting experiment when signal and background events obey Poisson statistics. The background events are estimated previously, either by Monte Carlo events, ``idle" run events or any other reasonable way. The here studied method applies only when the background and signal are completely independent events, i.e, they are non-coherent events. The profile   $\chi^2$-function has shown to have a fast convergence, with fewer events, to the ``true'' values for counting experiments as shown in MC toy tests. It works properly even when the bin contents are low and also when the signal to background ratio is small. Other interesting points are also presented and discussed. One of them is that the background parameter does not need to be estimated with very high precision even when there are few signal events during a fitting procedure.  An application to Higgs boson discovery is discussed using previously published ATLAS/LHC experiment data. 

\keywords{Maximum likelihood \and $\chi^{2}$-function \and parameter estimation \and Monte Carlo}
\end{abstract}

\section{Introduction}
\label{intro}
It has been proposed  many methods in order to get the best parameters from a fitted curve. The main motivation for that effort is the signal treatment with low statistics, when the standard least square method (LSM) gives non acceptable results since the distributions are non-Gaussian. Non-Gaussian distributions are important, for example, in particle physics, where the experimental physicists have frequently to deal with counting experiment with few events data sets. A modified LSM was developed by Phillips \cite{bi1} obtaining satisfactory results. Later Awaya \cite{bi2} published an approach to fit data sets with poor statistic without the use of $\chi^2$-function minimization. This later work uses the area preservation technique.\par
The importance of a $\chi^2$-function can be read in a very interesting paper by Baker and Cousins \cite{bi3}. They discuss some topics such as point estimation, confidence interval estimations, goodness-of-fit testing, biased estimation, etc., when fitting curves to histograms using $\chi^2$-functions.\par
More recently, some authors derived methods based on different $\chi^2$-functions built from the maximum likelihood ratio test theorem that show faster convergence to the true value than the LSM. These methods do not present the LSM problems when some of the bins have low or zero contents \cite{bi3,bi4}. Although all those methods have a good performance, it is still necessary to fit the background in order to extract information about the signal.  In many complex cases when one has background of unknown ``shape'', it is necessary to fit a polynomial of a reasonable degree. In addition to the signal parameters, there are  the background parameters and their respective errors.  All these parameters can be strongly correlated making the signal characteristics analysis and interpretation very complex.\par
In particular, data analysis at LHC experiments at CERN use extensive MC background studies in order to find new rare discoveries and adequate and sensible statistical methods should be used.\par  
In this work one considers a $\chi^2$-function obtained from the profile likelihood for signal
fitting without fitting the background once this background is estimated previously.  The method is based on the idea of profile likelihood detailed by \cite{Murphy} and discussed by \cite{bi5,Rolke2005}.  It is shown in the next 2 Sections the road map to obtain the profile likelihood function and its corresponding $\chi^2$-function. Section 4 shows the profile $\chi^2$-function expression to be used when fitting data and Section 5 presents some systematic results of a ``toy'' Monte Carlo experiment, and some other points are also discussed.  Based on Section 5 considerations, some $\chi^2$-functions, for well defined backgrounds, is obtained in Section 6. In Section 7 it is shown some applications. The conclusions are written in the last section.
In order to be scholarly informative the text is self-contained as possible. 

\section{Profile Likelihood}
Let us assume a counting experiment such that the signal and background events are completely independent and that both obey Poisson distributions.  The background events are first estimated using the Monte Carlo (MC) methods, running the experiment in "idle" mode or by any other technique. Suppose that during the experiment $k$ data events are  obtained  and $m$ background events were previously estimated using MC methods.  Since the number of previously estimated MC events can depend on computational resources, it is possible to generate $\tau$ relative samples, such that 
 \begin{equation}
 \tau=\cal L_{MC}/\cal L_{EXP}
\label{eq:Lumratio}
 \end{equation}  
 where $\cal L_{EXP}$ and $\cal L_{MC}$ are the experimental and MC integrated luminosities, respectively, as said in high energy physics jargon.  $\tau$ is the relative size of the MC background sample to the data sample and it is always larger than zero.  When one has very limited computer resources $0<\tau<1$.
Any information about the background is helpful in order to extract a better signal information as is shown in Section 5.
 The likelihood corresponding to the above discussed  case is

\begin{equation}
L(s,b;k,m,\tau) \propto (s+b)^{k}e^{-(s+b)}(\tau b)^{m}e^{-\tau b}
\label{eq:likel}
\end{equation}

\noindent 
where $s$ and $b$ are related to the signal and background distributions, respectively. As $\tau$ increases, our knowledge about the background parameter also increases and in the limiting case

\begin{eqnarray}
\lim_{\tau \longrightarrow \infty}(\tau b)^{m}e^{-\tau b} \propto \delta (b-\dfrac{m}{\tau})
  \end{eqnarray}
\noindent 
which means that the background parameter is exactly known.\\

There is an interesting heuristic way to eliminate the nuisance parameters related to $b$  and obtain a likelihood independent from $b$. It consists in finding the maximum likelihood estimator for background $\widehat{b}$ as a function of $s$ and replacing $b$ by $\widehat{b}$ in Eq. (\ref{eq:likel}). Taking the derivative of the Eq. (\ref{eq:likel}) with respect to $b$, one has 

\begin{equation}
\left.\dfrac{\partial}{\partial b}\:log\:L(s,b;k,m,\tau)\right|_{b=\widehat b} = 0.
\end{equation}

Solving the above equation and knowing that  $b\geq 0$, one gets

\begin{equation}
\widehat{b}(s) = \left(\dfrac{k+m-(1+\tau)s + {\Delta(s)}}{2(1+\tau)}\right)
\end{equation}
where
\begin{equation}
\Delta(s) =\sqrt{ \left[k+m-(1+\tau)s\right]^{2}+4 m (1+\tau)s} \ge 0
\end{equation}
\noindent
It is interesting to note that
\begin{equation}
\lim_{\tau \longrightarrow \infty} {\widehat b}(s) = m/\tau
\end{equation} 
\noindent and does not depend on $k$ and $s$.

Replacing $b$ by $\widehat b(s)$ in Eq. (\ref{eq:likel}) one obtains the profile likelihood $L_P(s;k,m,\tau)$, which does not depend on $b$. 
\begin{eqnarray}
L_P(s;k,m,\tau) \propto (s+\widehat b(s))^{k}e^{-(s+\widehat b(s))}(\tau \widehat b(s))^{m}e^{-\tau \widehat b(s)}
\end{eqnarray}

Solving the equation below, one gets the maximum of $L_P$ and the most probable value of $s$, $\widehat s$ 
 \begin{equation}
\left.\dfrac{\partial}{\partial s}\:log\:L_P(s;k,m,\tau)\right|_{s=\widehat s} = 0.
\end{equation}
The simple analytical solution of Eq.(9) is for

\begin{eqnarray}
\widehat s = max\left(0, k-\frac{m}{\tau}\right)
 \end{eqnarray}

\noindent
 since $s \geq 0$.  $\widehat s$ is just the maximum profile likelihood estimator of $s$.

\section{Profile $\chi^2_P$-function for Poisson signal and background} 

Let us construct now an approximate $\chi^2$-function from Eq.(8).
One  can construct the maximum profile likelihood ratio 

\begin{equation}
\lambda_P = \dfrac{L_P(s,k,m,\tau)}{L_P(\widehat{s},k,m,\tau)}
\end{equation}
where the denominator is the profile likelihood maximum. 
It occurs when $s=\widehat s$ as already shown in Section(2).  According to the 
maximum likelihood ratio theorem  \cite{bi6} one can construct now a profile $\chi^2$-function since 

\begin{equation}
 \chi_P^2 \approx -2\log \lambda_P
\end{equation}

\noindent
Eq.(13) below was written in such way to facilitate its implementation in a computer program

\begin{equation}
\begin{aligned}
 \chi_P^2 \, & = \, 2\left\{(s-\widehat s) + (\tau+1)\left(\widehat b(s)-\widehat b(\widehat s)\right) \right\} + \\
 & +2\left\{ k\ln\left(\frac{\widehat s +\widehat b(\widehat s)}{s+\widehat b(s)}\right)+ m\ln\left(\frac{\widehat b(\widehat s)}{\widehat b(s)}\right)\right\}
\end{aligned}
\end{equation} 
\newline
\noindent 
where $\widehat s$ and $\widehat b(s)$ are given by Eqs. (5) and (10), respectively, and so $\widehat b(\widehat s)$.

Analyzing the particular case of Eq.(5) when $m=0$, one has 
$$\widehat b(s)=max\left(0,\frac{k}{1+\tau}-s\right), $$
$$\widehat s=k$$ and the Eq.(13) above reduces to

\begin{equation}
\chi_P^2  = 2\left[s-k+(\tau+1) \widehat b(s) + \\
           k\left(\ln{k} -\ln{(s+\widehat b(s)})\right)\right]
\end{equation}

\noindent 
and in addition when $\widehat b(s)=0$, one obtains  Eq.(15), below, which is exactly the $\chi^2$-function obtained by \cite{bi3}, also mentioned in the PDG \cite{PDG}, when there is no background events
\begin{eqnarray}
 \chi_P^2=2\left[s-k+ k\left(\ln{(k)} -\ln{(s)}\right)\right]
\end{eqnarray} 
\par
\noindent 
So Eq.(15) is a particular result from the more general Eq.(13). 
It is shown in Fig.1 the $\chi^2_P$-function as a function of $s$ for $k=40$ and $m/\tau=10$ for 5 different $\tau$ values ranging from $0.1$ to $\infty$.
We notice that there is no large differences between $\tau=5$ and $\tau \longrightarrow \infty$ meaning that it is enough to have $\tau\approx5$, since $\tau \gg 5$ will be a waste of computer time, when the background is estimated by MC. Increasing previously our knowledge about the background also increase our knowledge about the signal since $\chi^2_P$-function becomes narrower up to a certain limit when $\tau\approx5$.  For $\tau\gg 5$ the signal statistical fluctuations dominates. We notice also that the main variation of  $\chi^2_P$-function with respect to $\tau$ occurs in the left region of its minimum, for Fig.1, $0 < s < 30$, then one has a significantly better lower limit for the parameter signal estimation
if one uses the $\chi^2$-function to find signal limits.\\

\begin{figure}
 \centering
 \includegraphics[width=0.52\textwidth]{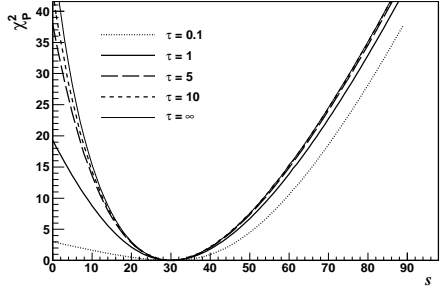}
 \caption{$\chi^2_P$-function as a function of s for different values of $\tau$ when $k=40$ and $m/\tau=10$}
 \label{Fig1}
\end{figure}

One can also obtain signal limits, $s_{min}$ and $s_{max}$, after solving the equation system below
for $s_{min}$ and $s_{max}$

\begin{eqnarray}
\left\{\begin{array}{ll}
\chi^2_P(s_{min})=\chi^2_P(\widehat s)+ \alpha \\
\\
\chi^2_P(s_{max})=\chi^2_P(\widehat s)+ \alpha \\
\\
0 \leq s_{min} < s_{max}
      \end{array}
\right.
\end{eqnarray}

\noindent 
\newline
The $\alpha$ value depends on the confidence level chosen.

\section{Fitting signal data with $\chi^2_P$-function}
Suppose one constructs a histogram with N bins labeled by the index $i$ running from 1 to N, with $\{k_{i}\}$ event data set and $\{m_{i}\}$ 
previously estimated background event set in the $\{i^{th}\}$ histogram bin set, respectively.
Let us suppose also that the background  was estimated previously using MC methods 
as already mentioned before.
Assuming that signal and background data are independent, the probability for the $i^{th}$ bin to have $k_{i}$ data events given $m_{i}$ previously estimated background events, both obeying Poisson distributions, is proportional to \cite{bi5}

\begin{equation}
\begin{aligned}
&L_{Pi}(s_i;k_i,m_i,\tau) \propto 
\left(s_i+\widehat b(s_i)\right)^{k_{i}} \times \\
 &\times exp\left\{-s_i-\widehat b(s_i)\right\}\left(\tau \widehat b(s_i)\right)^{m_{i}}exp\left\{-\tau \widehat b(s_i)\right\}
\end{aligned}
\end{equation}

\noindent
where $s_i$ is the signal distribution, $\widehat b(s_i)$ is a function of $s_i$  and $\tau$ is the ratio between the MC and  experimental luminosities.  Using the maximum likelihood ratio theorem for Eq.(17), one has
\begin{eqnarray}
\lambda_i =\dfrac{L_{Pi}(s_i;k_i,m_i,\tau)}{L_{Pi}(\widehat s_i;k_i,m_i,\tau)}
\end{eqnarray}

\begin{eqnarray}
\chi^2_P= -2 \ln\left(\prod_{i=1}^N \lambda_i \right)
\end{eqnarray}

\begin{eqnarray}
\chi^2_P=-2 \sum_{i=1}^N \ln{\lambda_i}
\end{eqnarray}

\begin{eqnarray}
\chi^2_P =\sum_{i=1}^N \chi_{P_i}^2
\end{eqnarray}

\noindent
where $\chi_{P_i}^2= -2 \ln\lambda_i$.
\bigskip
\noindent
The $\chi^2_P$-function will be minimized and it is given by the sum of all
$\chi^{2}_{P_i}$ that corresponds to N bin contributions,
where $s_{i}$ must be substituted by function $f(x_{i},\overrightarrow{\theta})$ that one wants to fit, being
$x_i$ the corresponding ordinate to $i^{th}$ bin and
$\overrightarrow{\theta}$  the parameter vector to be
fitted. The $\chi^{2}_{P_i}$ is given by

\par
\begin{equation}
\begin{aligned}
 \chi_{P_i}^2 &=2 \left(f(x_{i},\overrightarrow{\theta})-\widehat s_i\right) + \\
              &+ 2 (\tau+1)\left(\widehat b\left(f(x_{i},\overrightarrow{\theta})\right)-\widehat b(\widehat s_i)\right)  + \\
&+2 k_i\ln\left(\frac{\widehat s_i +\widehat b(\widehat s_i)}{f(x_{i},\overrightarrow{\theta})+\widehat b\left(f(x_{i},\overrightarrow{\theta})\right)}\right) +  &\\
&+ 2 m_i\ln\left(\frac{\widehat b(\widehat s_i)}{\widehat       b\left(f(x_{i},\overrightarrow{\theta})\right)}\right)
\end{aligned}
\end{equation}

\noindent 
where 
\begin{equation}
\widehat{b}(\widehat s_i) = \left(\dfrac{k_i+m_i-(1+\tau)\widehat s_i + {\Delta(\widehat s_i)}}{2(1+\tau)}\right)
\end{equation},

\begin{equation}
\Delta(\widehat s_i) =\sqrt{ \left[k_i+m_i-(1+\tau)\widehat s_i\right]^{2}+4 m_i (1+\tau)\widehat s_i} \ge 0
\end{equation}
and
\begin{equation}
\widehat s_i = max\left(0, k_i-\frac{m_i}{\tau}\right)
\end{equation}

Note that the Eqs.(22-25) depend just on $f(x_i,\vec \theta)$, $k_{i}$, $m_{i}$ and $\tau$. This is the great advantage of this method. We do not need to fit the background distribution, and the only  necessary information from the background is $\tau$ and the number of background events set $\{m_{i}\}$ estimated previously by MC. The $\chi^2_P$-function  has already incorporated the background statistical fluctuations. Besides reducing the number of fitted parameters, this approach also does not present problems when one has few or no event in one or more bins as can occurs in long tail data. Even the bins with $k_i=0$ 
and/or $m_i=0$ contributes to the $\chi^2_P$-function.  It is is only necessary to fit the signal function parameters which will allow us to obtain a much cleaner and less noisy analysis. This will affect in a positive way the signal parameter covariance matrix. 

\section{On $\tau$ Dependency Amid Fitting Processes}
In order to test the proposed method and how is the $\tau$ dependence behaviour, random numbers were generated using toy Monte Carlo experiments, where the number of signal entries $N_{s}$ changed from $20$ to $10^4$ distributed in a histogram of $100$ bins, in such way that for each fixed number of entries, it was generated $10^4$ sets of signal points
and fitted each of them. The ratio of the number of signal $N_s$ and  background events $N_{b}$ were kept constant and they were generated according to relation

\begin{equation}
\dfrac{N_s}{N_s+N_b}\:= \dfrac{1}{10},
\end{equation}
and for the ``previously'' estimated background events
\begin{equation}
N^{prev}_{b}\:= \tau \:N_{b},
\end{equation}
then the significance defined here as 

\begin{equation}
\dfrac{N_s}{\sqrt{N_s+N_b}}
\end{equation}

\noindent goes from $1.41$ for $N_s=20$ to $31.6$ for $N_s=10^4$.
The distributions used to generate the points were a Gaussian signal with an exponential background, similar to the one expected for the Higgs discovery at LHC.
The exponential is given by $e^{-x}$ and the Gaussian is given by

$$\exp\left\{-\dfrac{(x-\mu_0)^2}{2\sigma_0^2}\right\}.$$

\noindent
The values $\mu_0=1.2$ and $\sigma_0=0.2$ were used to generate the Gaussian signal distributions.
The fits were performed in the range $x \in (0,5)$ with $100$ bins and for six different values of
$\tau$ ranging from $0.1$ to $10$. For each fixed number of entries the average value of the fitted parameters $\mu$ and $\sigma$  as well as their fluctuations $\Delta\mu=\sqrt{<(\mu -\mu_0)^2>}$ and $\Delta\sigma=\sqrt{<(\sigma -\sigma_0)^2>}$  
were calculated with respect to their respective ``true'' values, $\mu_0$ and $\sigma_0$. Gaussian amplitudes were also fitted but not
plotted since we are generating non-normalized distributions. The
MERLIN optimization package \cite{bi7} was used for the minimization procedure.
\par
It is shown in Figs.2-5 how fast the parameters converge to the true value 
as $\tau$ and $s$ increase. This agrees with what we foresaw, since 
increasing the $\tau$ value means that we have a better
background knowledge and consequently a better signal information. It is observed meanwhile that between
$\tau = 5$ and  $\tau = 10$ the improvement is minimum as expected from Sec.(3), and
therefore one does not substantially improve the signal information when estimating the background with $\tau>5$.

\begin{figure}
 \centering
\includegraphics[width=0.52\textwidth]{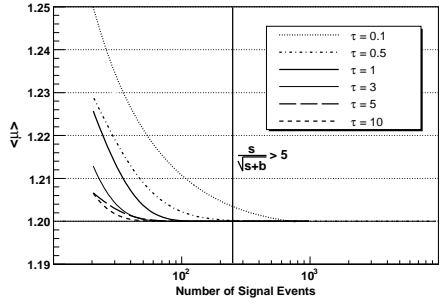}
 \caption{The average value of the fitted Gaussian parameter $\mu$ as a function of the number of signal events for different 
values of $\tau$. The region to the right of the vertical line corresponds to a significance greater than $5$.}
 \label{Fig2}
\end{figure}

\begin{figure}
 \centering
\includegraphics[width=0.52\textwidth]{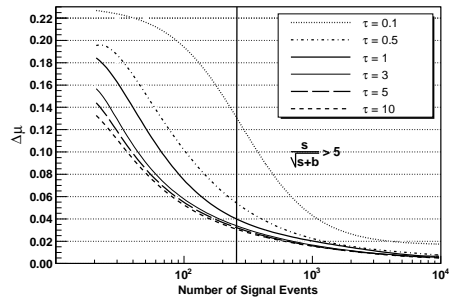}
 \caption{$\Delta\mu$  as function of signal events for different values of $\tau$. The region to the right of the vertical line corresponds to a significance greater than $5$.}
 \label{Fig3}
\end{figure}

\begin{figure}
 \centering
\includegraphics[width=0.52\textwidth]{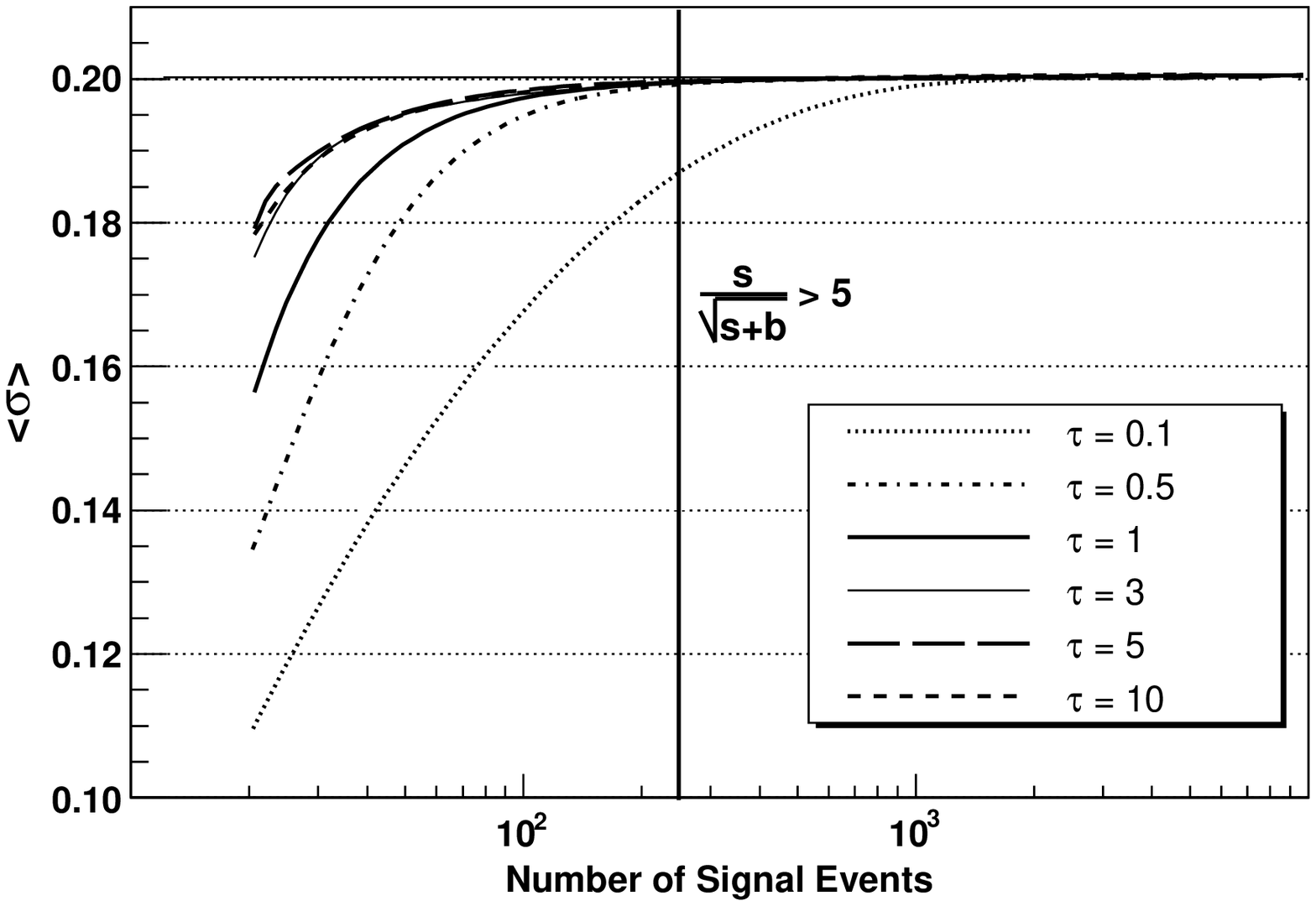}
 \caption{Same as Fig.2 but for the parameter $\sigma$.}
 \label{Fig4}
\end{figure}

\begin{figure}
 \centering
\includegraphics[width=0.52\textwidth]{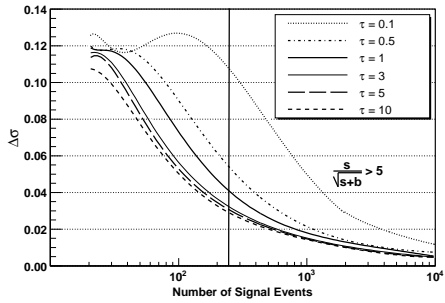}
 \caption{ Same as Fig.3 but for the parameter $\sigma$.}
 \label{Fig5}
\end{figure}

Comparing Fig.2 and Fig.4, it is interesting to note that while 
 $<\mu>$ has overestimated values for small number of signal  entries, 
$<\sigma>$ has underestimated values but both converge to the true value 
monotonically.
Note also that the bin width for our case is equal to $0.05$ and that both, $\Delta \mu$ and $\Delta \sigma$, reach this value when the significance $s/\sqrt{s+b} \approx 5$ for $\tau=1$, as shown in Fig.3 and Fig.5.  For $\tau>3$ and number of signal events  larger than $100$, $\Delta\mu$ and $\Delta\sigma$ are already smaller than the bin width.  This $100$ signal events corresponds to a significance around $3.16$.

Other signal distributions (Breit-Wigner and Moyal) and other background distributions (constant, straight line, exponential) were also studied and in all cases we obtained a similar behavior as shown in Figs 2-5. 
In all cases the average fitted parameters converges monotonically to the true value, with some starting with underestimated and others with overestimated values depending on the combination of signal and background distributions.

We also compared the proposed method with the LSM when the background was first fitted 
with previously generated background events of known distributions.
We notice that the here proposed approach presents a much faster
convergence to the true value when the number of events increases gradually and the fitted parameter fluctuations $\Delta\mu$ and $\Delta\sigma$ are also smaller than the LSM. 
Moreover, the proposed method presents a monotonically 
convergence as the number of signal entries increase, while the LSM
oscillates underestimating and overestimating the ``true'' parameter values.
As the number of signal events increase and consequently the significance, both methods coincide.

\section{Approximate $\chi^2$-Functions when $\tau \longrightarrow \infty $ }
It has been shown in previous section that during the fitting processes, there are small differences 
for any $\tau >5 $, for different fixed values of  $k$, the total number of events in a histogram bin.
Let us simplify the expression of Eq. (13) for the case $ \tau \longrightarrow \infty$. Calculating the limit of Eq(5)
when $ \tau \longrightarrow \infty$ one obtains Eq.(7), and this result does not depend neither of $s$ nor $k$.  
Let us define $\widehat B$ as 

\begin{equation}
\widehat B = \lim_{\tau \rightarrow \infty} \widehat b = \dfrac{m}{\tau}.  
\end{equation}

According to the profile likelihood approach, by replacing  
$\widehat b$ by $\widehat B$ in Eq.(2), one obtains a much simpler likelihood expression
\begin{equation}
L(s;k,\widehat B) \propto (s + \widehat B)^{k}e^{-(s +\widehat B)}
\end{equation}.

One can construct now a $\chi^2$-function from the above likelihood as described in Sec. (3):
\begin{equation}
\chi^2_{P(\tau > 5)}= 
2 (s-\widehat s) +2 k \ln \left(\dfrac{\widehat s + \widehat B}{s + \widehat B}\right), 
\end{equation}

\noindent where $\widehat s$ is now
\begin{equation}
\widehat s = max(0, k- \widehat B).   
\end{equation}

For multiple independent Poisson backgrounds, $\widehat B$ is, for all $\tau_i >> 5$, in Eq.(31)
\begin{equation}
\widehat B = \sum_{i=1}^{NBackg}\dfrac{m_i}{\tau_i}  
\end{equation}

It is interesting to note that using a Bayes approach, starting from Eq.(2), using Eq.(3) and a constant {\it a priori} background probability, one obtains the following likelihood
\begin{equation}
L(s,b;k,m,\tau) \propto (s+b)^{k}e^{-(s+b)}\delta\left(b-\dfrac{m}{\tau}\right),
\end{equation}
and integrating over $b$, in order to eliminate the nuisance parameters, one obtains the same Eq.(31). For large values of $\tau$, the profile and Bayes likelihood give the same results.   

Another approach is to follow Ref. [4], where a Poisson distribution is transformed in an approximated Gaussian distribution. The result for $\tau \rightarrow \infty$ is
\begin{equation}
\chi^2_{G(\tau > 5)}= 
2 (s-\widehat s) + (2 k +1) \log \left(\dfrac{2(\widehat s + \widehat B)+1}{2(s + \widehat B)+1}\right)
\end{equation}
The above expression is easy to handle in a computer, it takes in account even bins with contents equal to zero during the fitting processes and there are no singularities.

For large values of $k$ and $\tau > 5 $,  the Poisson distribution, Eq.(30), behaves like a Gaussian distribution with mean $\widehat s = max(0,k-\widehat B)$ and variance $\sigma=\sqrt{k}$, and one obtains
\begin{equation}
\chi^2_{LSM}= \dfrac{(s-\widehat s)^2}{k},
\end{equation}
\noindent 
which resembles the traditional least square method.

Fig.6 shows the results for the different $\chi^2$-functions when $k=20$, $m=50$, $\tau=5$ and $\widehat B = 10$.

\begin{figure}
\centering
\includegraphics[width=0.50\textwidth]{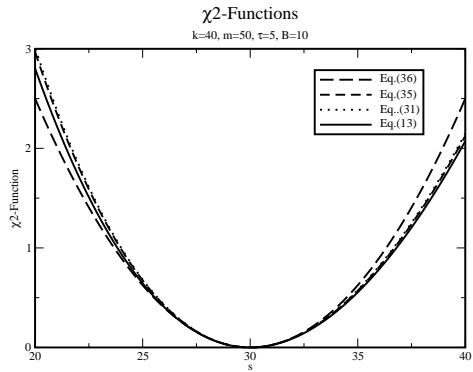}
\caption{Comparing the different $\chi^2$ functions. It is used $k=40$, $m=50$, $\tau=5$ for Eq.(13) and it is used $k=40$ and $\widehat B = 10$  for Eqs.(31,35,36) which correspond to $\tau \rightarrow \infty$.}
\label{Fig6}
\end{figure}

Eq.(36) will generate symmetrical errors while the others will generate asymmetrical errors with more restrictive lower bound errors than the upper bound errors. 

This section $\chi^2$-functions can be used to fit data after following Sec. (4). It is necessary to replace $s$ by $f(x_i, \overrightarrow{\theta})$ for each bin and then sum them up.     

\section{Applications}

Let us now see some use of profile likelihood and $\chi^2_P$-function with real data. We divide them in two types: one is the binned  when we know the shape of the signal distribution and the second when the signal and background shapes are unknown.

\subsection{Fitting the Higgs Mass}
\begin{figure}
 \centering
\includegraphics[width=0.52\textwidth]{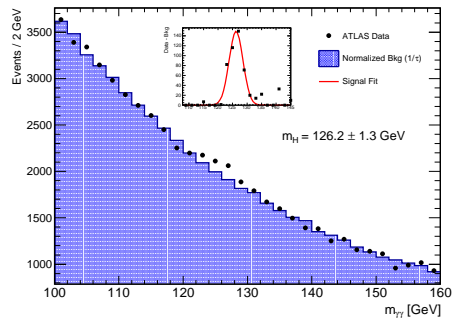}
\caption{Diphoton invariant mass distribution. The black point are the ATLAS data, and the full histogram is the normalized background previously estimated with $\tau = 5$. The red solid line is the fitted signal.}
 \label{Fig7}
\end{figure}

We apply the $\chi^2$-functions derived above to estimate the higgs boson mass from LHC data. The data considered are those from the ATLAS experiment that led to the higgs discovery in 2012 \cite{higgs}, with the higgs decaying in two photons. Figure \ref{Fig7} shows the diphotons invariant mass distribution from data and the estimated background. To perform the fit, the signal is modeled as a Gaussian and the background, for simplicity, is estimated using a toy MC with $\tau = 5$ and a fourth-order polynomial function. The signal fit resulted in a higgs mass of $m_H = 126.2 \pm 1.3$ GeV and width of $2.4 \pm 0.1$ GeV. The fitted higgs mass is in good agreement with the ATLAS measurement of $m_H = 126.0 \pm 0.4$ (stat) $\pm 0.4$ (sys) GeV. The error on the ATLAS measurement is smaller since it combines two different higgs decay channels.

\subsection{Subtracting Histograms and Obtaining Signal of Unknown Shape}
\begin{figure}
 \centering
\includegraphics[width=0.48\textwidth]{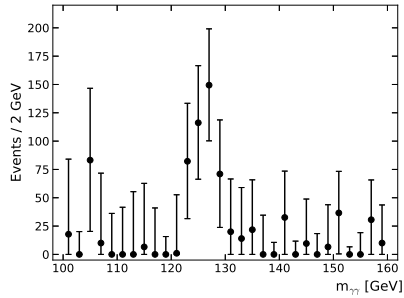}
 \caption{Signal extracted using a previously estimated background with $\tau = 5$. The dots correspond to $\widehat s = max(0,k-m/5$ and the limits on the signal in each bin correspond to $\chi^2_P(s)=1$ and $s>0$.}
 \label{Fig8}
\end{figure}

One can use the profile likelihood to extract signals of unknown shape in a binned distribution when the previously background distribution is known.  This is equivalent to subtract a background histogram from a data histogram in order to see if there is an excess of events in certain histogram region.
To illustrate the procedure, we consider the same data from the histogram of Fig. 7. The signal estimated for each bin is given by Eq.(25) and its error, in general asymmetric, is given by solving the system of Eq.(16) for a certain $\alpha$ value. The result for $\alpha = 1$ is shown in Figure \ref{Fig8}, from where we can see an excess of events mainly between  120  and 130 GeV. In order to quantify the result obtained we calculate the $p$-value under the hypothesis that one has no signal. Taking into account the bins between 123 and 130 GeV, one gets a $p$-value of $\sim10^{-4}$, corresponding to a signal significance of $\sim$ 3.5$\sigma$.

\section{Conclusions}

Usually the $\chi^{2}$ methods need the background curve to be
fitted before fitting the signal. The method here presented does
not require to fit the background so one does not need to
known its shape. The simulations with ``toy'' Monte Carlo shows that
the performance of this method is quite satisfactory with a fast
convergence to the parameter ``true'' values as $\tau$ and the number of signal events increases. Moreover, the method does
not present problems when the bin contents is low or even zero and the parameter estimators converge monotonically to the ``true'' values. We showed also that it is a waste of computer time to produce relative MC background samples for $\tau \gg 5$ and it is  possible to use the $\chi^2_P$-function to subtract background of unknown shape from data in order to obtain signal candidates or discovery also of unknown shape.  Although this analysis was done for data and background obeying Poisson distributions, it is possible to reproduce this study for different background and signal distributions. The $\chi^2_P$ (Eq.13) was implemented in Python and C++/ROOT. The codes are free available from the authors.


\end{document}